\documentclass[aps,prl,twocolumn,groupedaddress]{revtex4}
\usepackage{graphicx}
\begin{document}

\title{Holographic Entropy Packing inside a Black Hole}

\author{Aharon Davidson}
\email[Email: ]{davidson@bgu.ac.il}
\author{Ilya Gurwich}
\email[Email: ]{gurwichphys@gmail.com}

\affiliation{Physics Department, Ben-Gurion University of the Negev,
Beer-Sheva 84105, Israel}

\date{August 5, 2010}

\begin{abstract}
	If general relativity is spontaneously induced, the black
	hole limit is governed by a phase transition which occurs
	precisely at the would have been horizon.
	The exterior Schwarzschild solution then connects with a
	novel core of vanishing spatial volume.
	The Kruskal structure, admitting the exact Hawking imaginary
	time periodicity, is recovered, with the conic defect defused
	at the origin, rather than at the horizon.
	The entropy stored inside any interior sphere is
	universal, equal to a quarter of its surface area, thus locally
	saturating the 't Hooft-Susskind holographic bound.
	The associated Komar mass and material energy functions
	are non-singular.
\end{abstract}


\maketitle

Black hole thermodynamics is anchored to the well known
Bekenstein-Hawking area entropy formula \cite{HB}.
Intriguingly, neither the Gibbons-Hawking  \cite{GH} Euclidean
path integral derivation, nor the more locally oriented Wald's
\cite{Wald} derivation, make use of the black hole interior.
This seems to tell us that aside from the singularity in its
origin, the interior region of a black hole is quite a 'boring'
place.
Following this line, the area entropy has triggered the
fascinating idea that, from some obscure reason, no physical
degrees of freedom can reside within the interior of a black
hole. 
If it is the case, these degrees of freedom, whatever they
are, are owed to live on or near the horizon surface, with
one bit of information per a quarter of Planck area \cite{bit}.
The brane paradigm model, which introduces the notion
of a stretched horizon \cite{stretch}, is a realization of this
idea.
The catch is, however, that the horizon looks perfectly
innocent to in-falling matter.
The apparent inconsistency between the horizon as a physical
entity and as the mere point of no return has ignited a famous
debate in the physical society.
The black hole entropy formula has also inspired the so-called
holographic principle \cite{Hprinciple}.
The latter, primarily introduced by 't Hooft \cite{tHooft},
attempting to resolve the black hole information paradox,
and further developed by Susskind\cite{Susskind} to deal
with black hole complementarity,
is recently gaining theoretical support from the AdS/CFT
duality \cite{AdsCFT}.

Apparently, as far as entropy packing is concerned, the
interior of a black hole seems to be superfluous, certainly
within the framework of general relativity per se.
However, if general relativity is not a fundamental theory, but
rather a spontaneously induced theory of gravity, the
black hole limit has been shown \cite{Essay} to be governed
by a phase transition which occurs precisely at the would
have been horizon.
The fully recovered general relativistic exterior solution then
connects with a novel core of vanishing spatial volume.
The idea of horizon phase transition \cite{PhaseTransition}
is not new, and so is the notion of black stars and fuzzballs
\cite{BlackStar}.
In this paper,  after reviewing the fine details of the
underlying phase transition geometry \cite{Essay}, we show
how the black hole entropy is consistently packed inside the
whole interior region.
This is done in a universal manner, while locally saturating
the 't Hooft-Susskind holographic bound,
and is formulated by eqs.(\ref{Mass}, \ref{entropy}).

A simple theory which allows for a spontaneously induced
general relativity, is given by the action
\begin{equation}
	I=\int \left(-\frac{\varphi^{2}{\cal{R}}}{16\pi}
	-\frac{1}{16\pi a}\left( \varphi^{2}-\frac{1}{G}\right)^{2}+
	{\cal L}_{m}\right)\sqrt{-g}~d^{4}x ~.
	\label{action}
\end{equation}
The role of the Higgs potential is to allow the conformally
coupled Brans-Dicke \cite{BD} scalar field $\varphi(x)$ to
acquire its vacuum expectation value
$\displaystyle{\langle \varphi \rangle=G^{-1/2}}$ by virtue
of Zee mechanism \cite{Zee}, and in full analogy with the
electro/weak interactions.
Note that the theory is fully equivalent to $f(R)=R+\frac{1}{4}aR^2$
gravity,
with stability  $\acute{\text{a}}$ la Sotiriou-Faraoni \cite{f(R)}
guaranteed for $a>0$.
The value of $a$ can be made as small as necessary to be compatible
with Solar System tests.
An arbitrary scalar kinetic term can always be added
without affecting our main conclusions, but even in the absence
of such a term, the scalar field $\varphi(x)$ is dynamical.
The fact that the matter Lagrangian ${\cal L}_{m}$ does not
couple to the scalar field defines the Jordan, rather than Einstein,
frame to be the physical one
(our main results are nevertheless frame independent).
For the sake of the present paper, our interest lies with static
spherically symmetric vacuum solutions, with the standard line
element taking the form
\begin{equation}
	ds^{2}=-e^{\nu(r)}dt^{2}+
	e^{\lambda(r)}dr^{2}+r^{2}d\Omega^{2}~.
\end{equation}
The general relativistic $\displaystyle{\varphi(r)=G^{-1/2}}$
Schwarzschild solution, which we hereby tag with some
$\epsilon=0$ (related to the scalar charge), is accompanied
by a more general class of asymptotically flat $\epsilon\neq 0$
solutions.
The metric associated with the latter solution has been recently
derived and analyzed \cite{Essay}. 
Some details of the analysis are crucial for the sake of the present
paper, so a brief review is in order.

The large distance expansion of the generic solution is quite
conventional, see ref.(\cite{Essay}), with a scalar charge $Q_{s}$
factorizing a variety of Yukawa suppressed terms at a typical
length scale $\sqrt{aG}$, namely
\begin{eqnarray}\nonumber
	&& e^{\nu(r)}\simeq
	1-\frac{2GM}{r}+\frac{Q_{s}}{\sqrt{aG}} N(r)
	\frac{e^{-r/\sqrt{aG}}}{r^{GM/\sqrt{aG}}}~,\\
	&& e^{-\lambda(r)}\simeq
	1-\frac{2GM}{r}+\frac{Q_{s}}{\sqrt{aG}} L(r)
	\frac{e^{-r/\sqrt{aG}}}{r^{GM/\sqrt{aG}}}~,
	\\ \nonumber
	&& G\varphi^{2}(r) \simeq 
	1+\frac{Q_{s}}{r} F(r)
	\frac{e^{-r/\sqrt{aG}}}{r^{GM/\sqrt{aG}}}  ~,
	\label{Q}
\end{eqnarray}
where $F(r), N(r), L(r)= 1+{\cal O}(r^{-1})$.
Clearly, for $Q_{s}=0$, the Schwarzschild solution is fully
recovered, not just asymptotically.
However, with the focus on the $Q_{s}\rightarrow +0$ limit,
one can run a full scale numerical solution, starting from
the above asymptotic expressions, and already notice the
appearance of a phase transition near the would have been
horizon, at $r\simeq 2GM$.
In other words, \emph{representing a 'level crossing' effect (one
parameter family degenerates), the limit $Q_{s}\rightarrow 0$
does not reproduce
the $Q_{s}=0$ solution.}
In fact, this phase transition has already been verified
analytically \cite{Essay}, with the derived transition profile,
which connects the Schwarzschild exterior with the novel
interior in the vicinity of
$r\simeq 2GM$, given by
\begin{eqnarray} 
	e^{\nu (r)}\simeq \frac{\epsilon}{6}
	e^{-\sigma (r)}\left(\frac{6}{\epsilon}-
	e^{\sigma(r)}\right) ~,~
	e^{\lambda (r)}\simeq  e^{2\sigma (r)}e^{\nu (r)} ~,
	\quad&&\\
	r-2GM\simeq 2GM\left(e^{-\sigma (r)}+
	\frac{\epsilon}{6}\log\left(\frac{6}{\epsilon}
	e^{-\sigma (r)}-1\right)\right) ~, &&
	\nonumber
	\label{transition}
\end{eqnarray}
with $e^{\sigma (r)}$ serving as a parametric function.
It is quite possible that
the limit $\epsilon\rightarrow 0$ is physically unattainable,
and that one has to settle for a tiny yet finite $\epsilon$,
such that the invariant width of the transition profile
eq.(\ref{transition}) is of order Planck length,
highly reminding us of the 'stretched horizon' \cite{stretch}.
The transition profile is insensitive to the terms involving
the scalar potential, and exhibits a remarkable self similarity
 feature,
as expressed by the fact that $\epsilon\rightarrow
k\epsilon$ only causes scale changes
$e^{\lambda}\rightarrow k^{-1}e^{\lambda}$
and $r-2GM\rightarrow k(r-2GM)$.
The exact relation $\epsilon(Q_{s})$ is generally quite
complicated, and at this stage, was only obtained numerically
by plotting $\frac{1}{2}r(\nu^{\prime}+\lambda^{\prime})$.
The only analytic exception being the large $\sqrt{aG}$ case, for
which 
\begin{equation}
	\epsilon \simeq \frac{3Q_{s}}{2GM} ~.
\end{equation}
The forthcoming analysis, however, does not depend on
the exact value of $\epsilon$, but only on the
$\epsilon \rightarrow 0$ limit.

While the exterior Schwarzschild solution is asymptotically
recovered, which is an important feature by itself, it now
connects with a novel
interior solution.
This new solution, which is completely different from
Schwarzschild interior, is characterized by
(i) No $t \leftrightarrow r$ signature flip, (ii) Drastically
suppressed $e^{\nu(r),\lambda(r)}$, and (iii) Locally
varying Newton constant.
Altogether, the interior metric is well approximated by
\begin{equation}
	ds_{in}^{2} \simeq -\frac{\epsilon}{6}
	\left(\frac{r}{2GM}\right)^{\frac{6}{\epsilon}-
	4}dt^{2}+
	\frac{6}{\epsilon}
	\left(\frac{r}{2GM}\right)^{\frac{6}{\epsilon}-
	6+2\epsilon}dr^{2}+r^{2}d\Omega^{2} ~,
	\label{dsin}
\end{equation}
with the built-in phase transition scale manifest.
Such a short distance analytic behavior of the line
element expresses the fact that all terms in the field
equations which are factorized by tiny $e^{\lambda(r)}$
are practically negligible now, including in particular
the scalar potential terms.
The corresponding locally varying Newton constant,
defined as $\varphi^{-2}(r)$, being
\begin{equation}
	G_{in}(r) \simeq  G
	\left(\frac{r}{2GM} \right)^{2-\epsilon} ~.
\end{equation}
Associated with any concentric inner sphere of a finite
surface area $A(r)=4\pi r^{2}$ is the invariant spatial
volume
\begin{equation}
	V(r) \simeq 4\pi \sqrt{\frac{2\epsilon}{3}}
	\left(\frac{r}{2GM}\right)^{\frac{3}{\epsilon}}
	(2GM)^{3} ~.
	\label{V}
\end{equation}
The fact that $V(r)\rightarrow 0$ for any $r<2GM$,
as $\epsilon\rightarrow 0$, is the reason why, unlike in
any other macroscopic system, the
black hole entropy cannot be proportional to its volume.
This makes one wonder though how can physical degrees
of freedom actually reside within the interior core.

To gain some more insight into the inner metric, one may
verify the existence of a Kruskal-Szekeres structure, with the
corresponding scale function being
\begin{equation}
	S_{\omega}(r)\sim
	\left(\frac{r}{2GM}\right)^{\frac{6}{\epsilon}
	(1-4GM\omega)-4} ~,
\end{equation}
or define proper a distance
$\eta(r)\simeq \frac{2GM\sqrt{6\epsilon} }{3-2\epsilon+\epsilon^2}                                
\left(\frac{r}{2GM}\right)^{\frac{3}{\epsilon}-2+\epsilon}$,
and expand the inner metric up to ${\cal O}(\epsilon)$ pieces,
to expose the Rindler structure of the ${\Re}_{2}$
sub-metric
\begin{equation}
	ds_{in}^2\simeq-
	\frac{\eta^2 dt^2}{16G^2 M^2}+d\eta^2+
	4G^{2}M^{2}\left(\frac{3\eta^{2}}{8G^{2}M^{2}
	\epsilon}\right)^{\frac{\epsilon}{3}}
	d\Omega^2 ~.
	\label{Rindler}
\end{equation}
The recovery the exact Hawking's imaginary time periodicity
$\displaystyle{\Delta \tau =8\pi GM}$
is regarded as the anchor connecting us to black hole
thermodynamics.
Notice, however, that unlike in the original Schwarzschild
case, the Euclidean origin corresponds now to the center
of spherical symmetry $r=0$ rather than to $r=2GM$.

Inside the inner core, the Kretschmann scalar is well
approximated by
\begin{equation}
	K={\cal R}^{\mu\nu\lambda\sigma}
	{\cal R}_{\mu\nu\lambda\sigma}
	\simeq \frac{16\epsilon^{2}}{9\eta^{4}} ~,
\end{equation}
(and ${\cal R}^{\mu\nu} {\cal R}_{\mu\nu}\simeq
\frac{8\epsilon^{2}}{9\eta^{4}}$),
but also contains the apparently negligible  term $\frac{4}{r^4(\eta)}$.
Note that $K$ is suppressed by a factor $\frac{3}{4}\epsilon^{2}$
in comparison with its Schwarzschild
analogue, and consequently, the singularity analysis bifurcates.
(i) Clearly, for any finite $\eta$, as small as desired, the limit
$\epsilon\rightarrow 0$ is \emph{regular}.
The Rindler sub-metric gets then multiplied by a 2-sphere of
radius $2GM$, and consistently, the Kretschmann curvature
approaches the Schwarschild 'near horizon' value of $\frac{4}{(2GM)^{4}}$.
However, (ii) For any finite $\epsilon$, as small as desired,
the limit $\eta\rightarrow 0$ is \emph{singular}.
Whereas the pseudo-horizon does provide some protection
from the singularity (e.g. it takes an infinite amount of time
for light from the singularity to reach any external observer),
an observer willing to wait long enough will see unbounded
high curvature.
Such a behavior is far worse than that of the Schwarzschild
solution, and constitutes a severe problem.
It may be that a more complicated Lagrangian could alleviate this
behavior, or that quantum effects could cure it.
At any rate, invoking Hawking's analysis, and taking into account
the accompanying $\eta^4$ redshift of the radiation emanating
from the inner core, an external observer would encounter the
standard Hawking radiation.

At this point, we introduce the effective energy
density $\rho(r)$ and the effective pressure $p(r)$.
To clarify what exactly do we mean by the word effective,
consider a physicist convinced that general relativity is
the fundamental theory of gravity, and therefore totally
unaware of its hereby advocated spontaneously generated
nature.
Such a physicist would re-arrange the underlying field
equations into their basic Einstein form
${\cal R}_{\mu\nu}-\frac{1}{2}g_{\mu\nu}{\cal R}=
8\pi G{\cal T}_{\mu\nu}$, moving all terms, save for the
Einstein tensor itself, to the r.h.s., thereby
constructing the effective energy/momentum tensor
${\cal T}_{\mu\nu}$.
This does not change the fact that the physical metric
remains a Jordan (not Einstein) frame metric, as dictated
by the fact that in this frame, by construction, ${\cal L}_m$
does not couple to the scalar field.
Given the metric from eq.(\ref{dsin}), one can easily verify
that the effective energy/momentum tensor is
\begin{equation}
	{\cal T}^{\mu}_{~\nu} \simeq
	\frac{1}{r^{2}}\left( \frac{r}{2G M} \right)
	^{-\frac{6}{\epsilon}+6}
	Diag\left(-1,1,0,0\right)~.
	\label{T}
\end{equation}
Clearly, $\rho$ and $p$ diverge at the $\epsilon\rightarrow 0$
limit.
However, they do so in such a way that 
$e^{\frac{1}{2}(\nu+\lambda)}{\cal T}^{\mu}_{~\nu}$ actually
converges, a crucial observation for the forthcoming discussion.
 
Consider now the proper mass $M(r)$ associated with some
concentric sphere $\Omega(r)$.
Recalling the recovery of the exterior Schwarzschild geometry
for $\epsilon\rightarrow 0$, $M(r)$ must asymptotically
approach $M$ for all $r>2GM$.
Furthermore, $M(r)$ must be a finite monotonically increasing
function of the circumferential radius $r$.
Unfortunately, general relativity does not offer a unique definition
for the term mass.
Obviously, the naive mass formula $4\pi\int
\rho~r^{2} dr$ will not do  as it is non-covariant (and the
integrant diverges), and the ADM mass only makes sense
globally at asymptotically flat spatial infinity.
At this stage, the tenable candidates are:

\noindent (i) The Komar mass \cite{Komar} - Invoking
Stoke's theorem and performing the angular integration, it
acquires the form
\begin{equation}
	M_{K}(r)=\frac{1}{2G}r^{2}
	e^{\frac{1}{2}(\nu-\lambda)}\nu^{{\prime}} ~.
	\label{Komar}
\end{equation}
This mass function requires the presence of a timelike Killing
vector.

\noindent (ii) The material energy - Following Weinberg
\cite{Weinberg}, this is the integration of the energy as measured
in a locally inertial frame.
The naive mass formula is then supplemented by the missing
$\sqrt{-g_{tt}~g_{rr}}$ factor to give
\begin{equation}
	M_{W}(r)=\frac{1}{G} 
	\int_0^{r} e^{\frac{1}{2}(\nu+\lambda)}
	\left(  1+e^{-\lambda}(\lambda^{\prime}r-1)
	\right)dr ~.
	\label{Weinberg}
\end{equation}
Given the metric eq.(\ref{dsin}), and for
$\epsilon\rightarrow 0$, we obtain for the entire core the
mass function
\begin{equation}
	M(r) = M\left(\frac{r}{2GM}\right)^2 ~.
	\label{Mass}
\end{equation}
The two mass definitions eqs.(\ref{Komar},\ref{Weinberg})
are distinguished from each other only by their different
${\cal O}(\epsilon)$ corrections.
The striking feature now is that
unlike the Schwarzschild case, the present singularity does not
induce any mass contribution whatsoever.
The emerging Komar mass function appears to be well defined and
non-singular!

The main question now is how will the black hole
configuration change when supplementing $M$ by a tiny
amount $\delta M$?
First of all, as is well known, its overall radius would gain
a $2G\delta M$ increase.
But furthermore, appreciating the recovery of the Hawking
temperature $\displaystyle{T_{H}=(8\pi GM)^{-1}}$ from the
Rindler structure of the metric eq.(\ref{Rindler}), such a
process would induce an entropy increase of
$\displaystyle{\delta S=\frac{\delta M}{T_{H}(M)}}$,
leading eventually to the famous Bekenstein-Hawking formula
\begin{equation}
	S_{BH}=4 \pi GM^{2} ~.
\end{equation}
This, however, comes with no surprise.
After all, as mentioned earlier, the interior Schwarzschild
solution does not seem to play any role in the game.
But in the present case, facing a self-similar profile transition
into a novel interior core with no signature flip, it becomes
meaningful and crucial to ask what portion of the above
$S_{BH}$ entropy, if any, which we denote by $S(r)$, is stored
within an arbitrary inner sphere of a finite surface area
$4\pi r^{2}$ hosting the mass $M(r)$?

The most general change in $M(r)$, namely
\begin{equation}
	\delta M(r)=M(r) \left(
	2\frac{\delta r}{r}-\frac{\delta M}{M} \right) ~,
\end{equation}
takes into account both the trigger shift $M\rightarrow M+
\delta M$, as well as some suitable correction $\delta r$ at
 any arbitrary inner sphere radius $r$.
For the special case $r=2GM$,  for example, we have
$\delta r/r=\delta M/M$.
The crucial point to notice now is that, for
$\epsilon \rightarrow 0$, the volume $V(r)$ is solely a
function of  the ratio $r/M$, with $M$
being the only length scale in the inner metric.
In other words, the new ($M+\delta M$)-configuration held in
thermal equilibrium is nothing but a linearly stretched
version of the old $M$-configuration.
Thus, conveniently choosing $\delta r$, such that
$\delta (r/M)=0$, for all
$r<2GM$, will assure $\Delta V=0$.
Implementing this physical choice, the first law of
thermodynamics (with the Hawking temperature $T_{H}$ at
infinity) can be put to work in the following form
\begin{equation}
	\frac{\delta S(r)}{8\pi GM} =
	\delta M(r)=\frac{r\delta r}{4G^2 M}~.
\end{equation}
Taking advantage of the $M$ cancelation, we finally arrive at
our main result
\begin{equation}
	S(r)=\frac{\pi r^2}{G}=S_{BH}
	\left( \frac{r}{2GM} \right)^2 ~.
	\label{entropy}
\end{equation}
The emerging entropy packing profile turns out to be (i) Locally
holographic, i.e. exhibits proportionality to $A(r)$ for every
$r\leq 2GM$, and (ii) $M$-independent, and thus universal.
The overall picture is then of an onion-like model.
The entropy of any inner sphere is maximally packed, and
unaffected by the outer layers.
Any additional entropy is maximally packed on its own external
layer, with $M$ as well as $M(r)$ being adjusted accordingly.
In some sense, each layer acts as an event horizon \cite{Paddy},
with the local light cone $\frac{dr}{dt}\simeq\pm \frac{\epsilon}{6}
\left(\frac{r}{2GM}\right)$
getting closed at the $\epsilon\rightarrow 0$ limit for every
$r$ in the core.
Notice that the derived entropy function eq.(\ref{entropy}) is
in full agreement with the 't Hooft-Susskind holographic bound
\cite{tHooft,Susskind}. 
In fact, it locally saturates this bound!

An interesting point has to do with the entropy to energy ratio,
which in our case reads
\begin{equation}
	\frac{S(r)}{M(r)}=\frac{S_{BH}}{M} =4\pi GM \geq 2\pi r ~.
	\label{ratio}
\end{equation}
This is in apparent violation of Bekenstein's universal entropy
bound \cite{Bbound}.
The reason seems to be the following.
Whereas the entropy $S(r)$ of some inner sphere is universal,
the associated mass $M(r)\sim M^{-1}$ is affected by the total
mass of the system.
Admittedly,  Bekenstein's universal bound is relevant
\cite{Bviolation} only for weakly self gravitating isolated physical
systems (and for these it is a much stronger bound than the
holographic one), but it may still
be applicable here provided the radius appearing in universal
bound is of the entire system rather than of a subsystem.

Altogether, the emerging local realization of the holographic
principle, although highly unconventional, is very pleasing.
The new entropy packing profile sheds new light on the way
information is stored within a black hole, and this is achieved
without invoking string theory or the AdS/CFT correspondence.
Rather than envision bits of information evenly spread solely
on the horizon surface or in its vicinity, a bit per Planck area,
they are now universally and holographically spread in the
whole black hole interior.
Our results stem from the spontaneous generation of general
relativity, and are insensitive to the explicit form of the scalar
potential.
Crucial to our analysis is the identification of the tenable
mass definition, with the Komar mass and the material energy
being the leading candidates.
This choice is a posteriori justified once the holographic
structure is fully revealed.
Furthermore, the non-singular mass distribution eq.(\ref{Mass})
appears to be intimately related to, and thus as fundamental as,
the entropy distribution eq.(\ref{entropy}).
Exactly the same structure is expected to hold once the
cosmological constant and/or the electric charge enter the
game.

\acknowledgments{It is a pleasure to cordially thank Shimon
Rubin and Eduardo Guendelman for valuable discussions.
Special thanks to PRL's DAE Prof. Steven Carlip for insisting
on the clarification of the singularity structure.}


\section{Bibliography}
\begin{thebibliography} {}
\bibitem {HB}
	S.W. Hawking, Phys. Rev. Lett. 26, 1344 (1971);
	J.D. Bekenstein, Lett. Nuov. Cimento 4, 737 (1972);
	J.D. Bekenstein, Phys. Rev. D7, 2333 (1973);
	S.W. Hawking, Nature 248, 30 (1974);
	J.D. Bekenstein, Phys. Rev. D9, 3292 (1974);
	S.W. Hawking, Comm. Math. Phys. 43, 199 (1975).
\bibitem {GH}
	G.W. Gibbons and S.W. Hawking, Phys. Rev. D15, 2752 (1977).
\bibitem {Wald}
	R.M. Wald, Phys. Rev. D48, R3427 (1993);
	V. Iyer and R.M. Wald, Phys. Rev. D50, 846 (1994).
\bibitem{bit}	
	J. D. Bekenstein, Lett. Nuovo Cim. 11, 467 (1974);
	V. Mukhanov, Pis. Eksp. Teor. Fiz. 44, 50 (1986) [JETP Lett. 44, 63 (1986)];
	J.D. Bekenstein and V.F. Mukhanov, Phys. Lett. B360, 7 (1995).	
\bibitem {stretch}
	K.S. Thorne, W.H. Zurek and R.H. Price, in \textit{Black Holes: The
	Membrane Paradigm}, p. 280 (Yale University Press, New Haven, 1986);
	L. Susskind, L. Thorlacius and J. Uglum, Phys. Rev. D48, 3743 (1993).
\bibitem {Hprinciple}
	D. Bigatti and L. Susskind, "Strings, branes and gravity" (Boulder),
	883 (1999);
	R. Bousso, Rev. Mod. Phys. 74, 825 (2002).
\bibitem {tHooft}
	G. 't Hooft, in \textit{Salam festschrifft} A. Aly, J. Ellis, and
	S. Randjbar Daemi eds, (World Scientific, 1993), [arXiv gr-qc/9310026];
	L. Susskind, J. Math. Phys. 36, 6377 (1995);
\bibitem {Susskind}
	L. Susskind, Jour. Math. Phys. 36, 6377 (1995);	
\bibitem {AdsCFT}
	J.M. Maldacena, Adv. Theor. Math. Phys. 2, 231 (1998);
	ibid. Int. Jour. Theor. Phys. 38, 1113 (1999).
\bibitem {Essay}
	A. Davidson and I. Gurwich,  Int. Jour. Mod. Phys. D19, 2345 (2010);
	\textit{Honorable mention}, Gravity Research Foundation (2010).
\bibitem {PhaseTransition}
	G. Chapline, E. Hohlfeld, R.B. Laughlin and D.I. Santiago,
	Int. Jour. Mod. Phys. A18,3587 (2003).
\bibitem {BlackStar}
	P.O. Mazur and E. Mottola,
	Proc. Nat. Acad. Sci. (PNAS) 101, 9545 (2004);
	C. Barcelo, S. Liberati, S. Sonego and M. Visser,
	Phys. Rev. D77, 044032 (2008);
	K. Skenderis and M. Taylor, Phys. Rep. 467, 117 (2008).
\bibitem {BD}
	C.H. Brans and R.H. Dicke, Phys. Rev. 124, 925 (1961).
\bibitem {Zee}
	A. Zee, Phys. Rev. D23, 858 (1981).
\bibitem {f(R)}
	A.A. Starobinsky, Phys. Lett. B, 91 (1980);
	For recent reviews see:
	T.P. Sotiriou and V. Faraoni, Rev. Mod. Phys. 82, 451 (2010);
	A. De Felice and S. Tsujikawa, Living Rev. Relativity 13, 3  (2010);	
\bibitem {Komar}
	A. Komar, Phys. Rev. 113, 934 (1959).
\bibitem {Weinberg}
	S. Weinberg, in \textit{Gravitation and Cosmology}, (Wiley, 1972),
	pp. 302-303.
\bibitem {Paddy}
	T. Padmanabhan, Class. Quant. Grav. 21, 4485 (2004).
\bibitem {Bbound}
	J.D. Bekenstein, Phys. Rev. D23, 287 (1981).
\bibitem {Bviolation}
	J.D. Bekenstein, Found. Phys. 35, 1805 (2005).
\end{thebibliography}
\end{document}